\documentclass[10pt,letterpaper]{article}
\usepackage[top=0.85in,left=2.75in,footskip=0.75in]{geometry}

\usepackage{amsmath,amssymb}

\usepackage{changepage}

\usepackage[utf8x]{inputenc}

\usepackage{textcomp,marvosym}

\usepackage{cite}

\usepackage{nameref,hyperref}

\usepackage{microtype}
\DisableLigatures[f]{encoding = *, family = * }

\usepackage[table]{xcolor}

\usepackage{array}

\newcolumntype{+}{!{\vrule width 2pt}}

\newlength\savedwidth

\raggedright
\usepackage[aboveskip=1pt,labelfont=bf,labelsep=period,justification=raggedright,singlelinecheck=off]{caption}

\makeatletter
\renewcommand{\@biblabel}[1]{\quad#1.}
\makeatother

\usepackage{lastpage,fancyhdr,graphicx}
\usepackage{epstopdf}
\fancyheadoffset[L]{2.25in}
\fancyfootoffset[L]{2.25in}
\begin{document}
	\vspace*{0.2in}
	
	\begin{flushleft}
		{\Large
			\textbf\newline{Real-time noise cancellation with Deep Learning} 
		}
		\newline
		\\
		Bernd Porr\textsuperscript{1\Yinyang*},
		Sama Daryanavard\textsuperscript{1\Yinyang},
		Luc\'ia Mu\~noz Bohollo\textsuperscript{1\Yinyang},
		Henry Cowan\textsuperscript{1\Yinyang},
		Ravinder~Dahiya\textsuperscript{2\ddag}
		\\
		\bigskip
		\textbf{1} Biomedical Engineering, James Watt School of Engineering, University of Glasgow, G12 8QQ, UK.
		\\
		\textbf{2} Bendable Electronics and Sensing Technologies (BEST) group, James Watt School of Engineering, University of Glasgow, G12 8QQ, UK.
		\\
		\bigskip
		
		%
		%
		\Yinyang These authors contributed equally to this work.
		
		\ddag These authors also contributed equally to this work.
		
		* bernd.porr@glasgow.ac.uk
		
	\end{flushleft}
	\section*{Abstract}
	Biological measurements are often contaminated with large amounts of
	non-stationary noise which require effective noise reduction
	techniques. We present a new real-time deep learning algorithm which
	produces adaptively a signal opposing the noise so that destructive
	interference occurs. As a proof of concept, we demonstrate the
	algorithm's performance by reducing electromyogram noise in
	electroencephalograms with the usage of a custom, flexible, 3D-printed, compound electrode. With this setup, an average of 4dB and
	a maximum of 10dB improvement of the signal-to-noise ratio of the EEG
	was achieved by removing wide band muscle noise. This concept has the
	potential to not only adaptively improve the signal-to-noise ratio of EEG but can
	be applied to a wide range of biological, industrial and consumer
	applications such as industrial sensing or noise cancelling
	headphones.
	
	
	\section*{Introduction}
	Low signal-to-noise ratios (SNR) exist in many application domains,
	such as communications, acoustics or biomedical engineering. In
	particular, the Electroencephalogram (EEG)
	\cite{Green1985,henry2006electroencephalography,Britton2016} has a
	low SNR ratio because of its low amplitudes, in the range
	of a few $\mu V$, which are contaminated by numerous sources, often
	orders of magnitude larger than the EEG signal itself
	\cite{fatourechi2007emg}. In this work, we target EEG as an example
	application and remove non-stationary electromyogram (EMG) noise. However, this
	concept of algorithmic SNR enhancement is not limited to this particular use case.
	
	There are two categorical approaches to increasing the SNR of an EEG signal: real-time processing
	and offline post-processing. Concerning the latter, by far the most popular
	approach is principal component analysis (PCA) or independent
	component analysis (ICA)
	\cite{bell1995information,Makeig1995,McMenamin2010,Fitzgibbon2007,Delorme2007}.
	PCA and ICA methods pre-analyse the raw signals in order to identify and separate the signal and noise components.
	This analysis is offline, requires the signal and noise
	relationships to be constant over time, and demands high
	computational power.
	
	Real-time algorithms, on the other hand, filter the EEG signals as they
	arrive, sample by sample, and do not rely on offline pre-analysis, for
	example, bandpass filters, the short time Fourier Transform or wavelet
	transform
	\cite{Ahmadi2012,jirayucharoensak2013online,Jirayucharoensak2019}.
	These techniques still require prior knowledge of the noise to
	tune the filter parameters. However, muscle noise is non-stationary
	due to both voluntary and involuntary contractions of surrounding facial
	muscles. A solution to this problem is real-time adaptive filtering in
	which the noise is removed by an adaptive algorithm
	\cite{Widrow1975,kher2016adaptive,he2004removal}.
	In cases where EEG
	electrodes are placed on top of the head (i.e. around $C_z$), one can
	assume that noise polluting the EEG originates from further afield and
	affects all electrodes in equal measure, while the EEG signals
	originate locally \cite{Fitzgibbon2015}. A second auxiliary electrode
	can be used for measuring the noise solely, this can then be
	\textsl{subtracted} from the main EEG electrode signal. The most
	popular design for such an auxiliary electrode is a ring-shaped
	electrode around the main EEG electrode where the noise is simply
	subtracted, this is called the ``Laplace operator''
	\cite{makeyev_ding_besio_2016,Fitzgibbon2015,garcia-casado2019,aghaei-lasboo2020,besio2006}.
	While the idea of simply subtracting the noise is perfect in theory,
	in practice, the relationship between the EEG generated in the brain
	and the resulting signals at the electrodes are complex and
	dynamic. This calls for a smart, compound electrode that implements an
	adaptive filter to continuously learn about the changing signal and
	noise conditions.
	
	In this paper, we present a proof of concept for a
	novel, compound electrode which is inexpensive and readily manufacturable, in
	combination with a new deep learning algorithm. This system adaptively
	removes the noise from the EEG by algorithmically creating an opposing signal to the
	noise which is, in turn, used to cancel out the noise. This is demonstrated
	below by the removal of wideband muscle (EMG) noise.
	
	\begin{figure}[!ht]
		\centering
		\includegraphics[width=\linewidth]{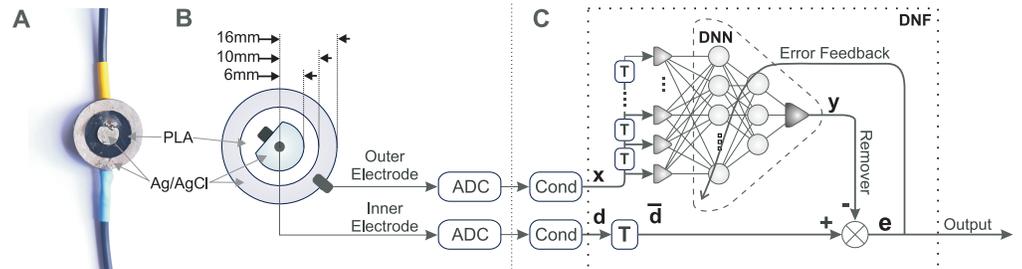}
		\caption{{\bf Electrode and Deep Neural Filter.} A: Photo of the manufactured compound electrode. The
			top wire (yellow) connects to the inner electrode and the bottom wire (blue) to
			the outer ring electrode. B: Top schematic view of the new compound
			electrode showing the inner electrode and the outer ring
			electrode. C: Signal processing of the two signals originating
			from the inner and outer ring electrodes: ADC = Analogue
			Digital Converter, Cond = standard signal conditioning such
			as high-pass filtering and 50~Hz removal. T = time delay.\label{fig:fig1_diagram}}
	\end{figure}
	
	\section*{Methods}
	\subsection*{General signal requirements}
	Let us consider a signal $\tilde{d}[n]$ measured with an ordinary electrode placed
	on the head of a subject:
	\begin{equation}
		\tilde{d}[n] = \underbrace{b[n] + m[n]}_{r[n]} + c[n]
	\end{equation}
	which is a superimposition of three signals:
	\begin{enumerate}
		\item $c[n]$ is the signal of interest generated by a stimulus or
		voluntarily, for example, in the setting of a brain-computer interface (BCI),
		\item $b[n]$ is the background
		EEG activity which is involuntary and unaffected by the stimulus, and
		\item $m[n]$ is the accumulation of all artefacts, in particular muscle activity (EMG).
		The latter two signals form the total baseline noise $r[n]$
		contaminating the EEG component $c[n]$ which is of interest for diagnostics or BCI applications.
	\end{enumerate}
	The task is now to reduce $r[n]$ as much as possible with the help of
	an opposing signal which ideally eliminates the noise from $\tilde{d}[n]$. As
	outlined in the introduction, we assume that the EEG originates
	\textsl{locally} from a small surface area of the head and that
	artefacts originate further afield and, therefore, they have a
	\textsl{global} and uniform strength across the scalp of the
	subject.
	
	The use of a second and linearly independent measurement would provide
	more information about the relationship between the global noise and
	the local EEG signal. Consequently, a compound electrode
	(Fig.~\ref{fig:fig1_diagram}A, B) is designed with the addition of an
	annular ring-like outer electrode around the inner electrode, which
	acts as the noise reference. Thus, the compound electrode collects two
	separate signals:
	\begin{eqnarray}\label{eq:alpha}
		\tilde{d}[n] = & r[n] + c[n] \quad & \textrm{Inner electrode: signal + noise} \label{inner}\\
		\tilde{x}[n] = & h[n] * \left(r[n] + \alpha \cdot c[n]\right) & \textrm{Outer ring electrode: noise reference} \label{outer}
	\end{eqnarray}
	where $0 < \alpha \ll 1$ models the crosstalk between the inner $\tilde{d}[n]$
	and outer $\tilde{x}[n]$ electrode signals, as the signal $c[n]$ of the inner
	electrode $d[n]$ will also stray into the outer ring. The noise in
	turn should ideally be present at both the inner part of the
	electrode $\tilde{d}[n]$ and the outer, ring electrode $\tilde{x}[n]$ but in
	practice, it will be a filtered version and is modelled with the filter
	$h[n]$. The goal of the learning algorithm is to render the signal
	from the inner electrode $\tilde{d}[n]$ as noise-free as possible so that
	ideally only $c[n]$ remains. Naively, one could simply subtract
	the outer electrode signal $\tilde{x}[n]$ from the inner one $\tilde{d}[n]$ to obtain
	a noise-free EEG but in practice, this is not possible because of
	changing noise-characteristics which are modelled here
	with the filter $h[n]$. Instead, we present a new machine learning algorithm which
	learns in real-time (i.e., when the data is being collected) to alter the signal
        from the outer noise reference electrode
	$\tilde{x}[n]$ in such a way that it eliminates the noise from the inner electrode
	which then results in a noise-free EEG signal. In the next two
	sections, we describe the electrode and the deep neural
	filter algorithm, respectively.

	\subsection*{Fabrication of the compound electrode}
	To record both the noisy EEG and a noise reference, a new compound
	electrode was designed (Fig.~\ref{fig:fig1_diagram}A, B). The physical
	design of the electrode was driven by durability, ease of manufacture
	and reliability. Polylactate acid (PLA) was chosen as the electrode
	material due to its compatibility, flexibility, and adhesive nature to
	silver/silver-chloride (Ag/AgCl) ink \cite{Rohaizad2019}. Ag/AgCl
	paste was selected for the conductive portion.
	Ag/AgCl was selected over alternative materials such as gold or stainless
	steel as it is conformable (in ink form) allowing easy application to the
	PLA and inexpensive. In addition, Ag/AgCl has a low half-cell voltage
	\cite{Lee2008}, meaning any oxidisation of the electrode will have a
	minimal effect on the sensitivity of the electrode.
	The combination of a
	flexible backing with conductive paste versus conventional, rigid,
	and often uncomfortable gold/platinum electrodes
	\cite{Flumeri2019,Lopez2014}, is advantageous as it allows for
	optimal skin-electrode contact. This optimal contact ensures minimal
	inter-electrode impedance resulting in an increased SNR for that
	electrode whilst also providing more comfort to the patient
	\cite{Suarez2018}, making long-term monitoring applications more
	viable. The compound electrode consists of two raised ring portions
	separated by a channel. The PLA geometry was 3D printed and the
	surface areas of the different electrode compounds were:
	\begin{equation}
		\begin{split}
			A_\textrm{Inner Ring} &= \pi \cdot {(6~\mathrm{mm})}^2 = 113~\mathrm{mm}^2\\
			A_\textrm{Outside} &= \pi \cdot {(16~\mathrm{mm})}^2 = 804~\mathrm{mm}^2\\
			A_\textrm{Inside} &= \pi \cdot {(10~\mathrm{mm})}^2 = 314~\mathrm{mm}^2\\
			A_\textrm{Outer Ring} &= A_\textrm{Outside} - A_\textrm{Inside} \\
			A_\textrm{Outer Ring} &\approx 490 ~\mathrm{mm}^2
		\end{split}
	\end{equation}
	When selecting the optimal surface areas, there is always
	a trade-off between localisation and signal strength.
	Increasing the area results in more contact
	area and thus receiving a stronger signal \cite{TANKISI2020243},
	but decreases the spatial resolution of the signal. In
	our application, we assume that EEG has a narrow spatial
	localisation and therefore requires a small surface area. We
	also assume that the noise has a broad spatial localisation
	as it is predominantly EMG artefacts perturbing
	the scalp all-across. An increase in the outer electrode area would,
	in theory, allow us to capture more EMG-noise for the
	algorithm to self-tune, however, as the signal strength
	is already orders of magnitude lower than the noise any
	realistic adaptation of surface area (given the necessity of
	comfort and localisation) would most likely result in
	negligible SNR enhancements.
	
	A layer Ag/AgCl paste was deposited on each of the raised rings using
	a plastic spatula. The Ag/AgCl was then cured at 70\textdegree{}C for
	1 hour. Fig.~\ref{fig:fig1_diagram}A shows the final printed
	electrode with Ag/AgCl applied. Two wires were connected to each electrode
	substrate by melting the copper onto the flexible PLA geometry using a
	soldering iron. Next, pure silver paste and epoxy were applied to the contact point
	to ensure reliable electrical contact and solidify the connection,
	respectively.	
	This electrode
	has proven to be robust and easy to both manufacture and integrate
	into a headband or EEG cap as a wearable device.
	
\subsection*{Experimental setup for EEG recording}
	Ethical approval for this experiment was obtained from the
        ethics committee at the Institute of Neuroscience and
        Psychology, School of Psychology at the University of Glasgow,
        reference number 300210055. In total, 20 subjects were
        recruited. Subjects were instructed to read an information
        sheet detailing the experiments and were permitted to
        participate after providing written consent. Every
        participant signed two copies of the consent form, one for
        the investigator and another for the
        participant to keep. The ethical approval letter, the
        information sheet and the consent forms are bundled together
        with the open-access dataset \cite{Bohollo2022}. The data was
        acquired using a two-channel data acquisition device
        (``Attys'', \url{www.attys.tech}) with the data acquisition
        programs \texttt{attys-ep} and \texttt{attys-scope}. Referring to the
        international 10-20 system, our compound electrode (see
        Fig.~\ref{fig:fig1_diagram}A, B) was placed on the subject's
        head at $C_z$, with its \textsl{inner} part connected to the
        positive input of Channel~1 and its \textsl{outer} ring
        electrode to the positive input of Channel~2 of the Attys. The
        A2 electrode (standard adhesive electrode behind right ear)
        was connected to the negative input of Channel 1, and the A1
        electrode (standard adhesive electrode behind left ear) was
        connected to the negative input of Channel 2 which also acted
        as ground. In the remainder of the paper we will just refer to
        the ``inner'' part and ``outer'' ring of the compound
        electrode and their corresponding signals (see
        Eqs.~\ref{inner} \& \ref{outer}). Each subject held two
        sessions with no intervals to guarantee consistent electrode
        signals:
	\begin{enumerate}
		\item\label{sess1}
		In this session the subject generates
		\textsl{EEG polluted with EMG noise}. To achieve this the subject was asked to
		contract their jaw muscle every 15~secs for two minutes to generate EMG
		noise. The sampling rate was $f_s = 500~\mathrm{Hz}$ and was chosen to
		obtain a flat response in the EEG frequency band of $0 \ldots 100~\mathrm{Hz}$
		due to the sigma-delta converter's smooth roll-off towards
		the Nyquist frequency of $250~\mathrm{Hz}$.
		\item\label{sess2} This session is used to obtain the \textsl{signal
			power of a noise-free EEG}. Since the subject cannot be paralysed to
		obtain EMG-free EEG signals, evoked potentials have been chosen to
		average out EMG noise. P300 visually induced oddball stimuli were
		used to determine the pure EEG signal $c[n]$ and its power (5~minutes). A
		black and white chequerboard inversion was presented every second
		and brightly-coloured horizontal bars as oddballs were randomly interspersed
		every 7~sec to 13~sec. The subject had the task of silently counting the
		number of oddball stimuli. The sampling rate was $f_s = 250~\mathrm{Hz}$; since
		the evoked potentials have low-frequency components,
		only their peak power is of interest in this work.
	\end{enumerate}
	We are now going to describe our new adaptive noise reduction algorithm
	which was then used to remove the EMG noise from the recordings of the
	different subjects.
	
\subsection*{Deep Neural Filter (DNF)}
	Fig.~\ref{fig:fig1_diagram}C shows the block diagram of our Deep
	Neural Network (DNN) which in conjunction with the additional
	building blocks becomes our novel Deep Neural \textsl{Filter} (DNF)
	to remove noise (see \cite{bernd_porr_2022_7100537} for the source
	code). Recall that the deep network exploits the assumption that the
	signal from the outer electrode $x[n]$ ideally just contains the
	noise and that the DNN learns to subtract it from
	the signal $d[n]$ originating from the inner electrode at the
	summation node ``X''.
	
	The error signal $e[n]$ of the network is also the final
        output of the DNF as is the case with LMS noise cancellation
        frameworks. This might appear counter-intuitive, as in
        classical applications of neural networks, the error $e[n]$ is
        expected to converge to zero.  At the same time, for filtering
        applications, the output is expected to be the clean signal.
        The key to resolving what appears to be a contradiction is to
        realise that before learning the output of the DNF $e[n]$ is a
        superposition of both EMG-noise and the pure EEG-signal. The
        noise component is expected to converge to zero through
        learning, leaving only the clean EEG-signal available at the
        output.  This is possible because the learning that takes
        place within the DNF network is not solely driven by the error
        feedback $e[n]$, rather, it is driven by the correlation of
        the error feedback $e[n]$ and the noise reference $x[n]$. If
        these two signals correlate, meaning some components of the
        noise is present at the output of the DNF, these shared
        components will be removed by the remover $y[n]$. This process
        will continue until the error feedback $e[n]$ and the noise
        reference $x[n]$ no longer correlate, meaning no components of
        the EMG-noise have remained at the output of DNF. This marks
        successful learning. In other words, the noise component of
        the output $e[n]$ has converged to zero despite it being a
        non-zero signal, this remaining component is the clean signal.
        In practice, the noise reference $x[n]$ often contains a
        certain amount of the pure EEG signal $c[n]$ which results in
        a reduction of the EEG signal at the DNF output. On the other
        hand, any uncorrelated noise between inner and outer
        electrodes such as thermal noise (approx. $65~\mathrm{nV} =
        \sqrt{1.380649 10^{-23}~\mathrm{J/K} \cdot 4 \cdot 310~\mathrm{K} \cdot
          1~\mathrm{K}\Omega \cdot 250~\mathrm{Hz}}$) or ADC-converter noise
        (approx. $100~\mathrm{nV}$) has no impact on learning and passes
        through the DNF.

	As outlined above the goal is to reduce EMG noise. However,
	eye-blink artefacts and slowly changing electrode drift have
	much higher noise power than EMG. Given that we are
	interested in EMG, we need to provide the reference noise input
	$x[n]$ with the muscle noise spectrum and remove the much
	more powerful low-frequency artefacts such as eye movement or
	baseline wander. We do this by employing a high-pass filter
	which captures the typical EMG spectrum which is flat above
	20~Hz but slowly decays in power below 20~Hz
	\cite{Whitham2007} (see also Fig.~AC in S1 Appendix).
        To force the DNF to learn the noise features of the
	EMG and not those of the EOG we set the 2nd order Butterworth
	high-pass for the noise reference $x[n]$ to $f_{c_x}=5~\mathrm{Hz}$
	which gives a shallow rise in the passband. The high-pass
	filter frequency for the inner signal $d[n]$ is not critical
	and was set to $f_{c_d} = 0.5~\mathrm{Hz}$ to simply remove the DC from
	the DC-coupled ADC converter so that all signals are DC-free:
	\begin{eqnarray}
		d[n] & = & \gamma \cdot \mathrm{HP}_{f_{c_d}}[n] * \mathrm{BS}[n] * \mathrm{LP}_\mathrm{ADC} * \tilde{d}[n] \\
		x[n] & = & \gamma \cdot \mathrm{HP}_{f_{c_x}}[n] * \mathrm{BS}[n] * \mathrm{LP}_\mathrm{ADC} * \tilde{x}[n]
	\end{eqnarray}
	where $\mathrm{HP}_{f_{c_d}}[n]$ and $\mathrm{HP}_{f_{c_x}}[n]$ are
	the $2^{nd}$ order high-pass Butterworth filters for the inner and outer
	electrodes, respectively. $\mathrm{BS}[n]$ is a $2^{nd}$ order Butterworth
	notch filter against powerline interference at 50~Hz.
	$\mathrm{LP}_\mathrm{ADC}$ is the low-pass characteristic of the sigma-delta converter with a cutoff at about half the sampling rate. The
	gain was set to $\gamma = 1000$ so that each neuron in the input layer
	of the neural network received values of approximately $\pm 0.2$V.
        The DNF uses as activation function $\tanh$ which saturates for values above
	approximately one but the input range of $x[n]$ at $\pm 0.2$
	will steer clear of any hard saturation. This also prevents vanishing gradients as the derivative of $\tanh$
        will be close to one in this regime and will be far off from becoming zero which only happens when $\tanh$ saturates.
        On the other hand at $0.2$ the $\tanh$
	is in its non-linear regime and the network will use its non-linear properties.
	
	Inspired by a Finite Impulse Response (FIR) filter, we send the signal
	of the outer electrode $x[n]$ through a tapped delay line with
	\begin{equation}
		N_{\mathrm{taps}_x} = \frac{f_s}{f_{c_x}}
	\end{equation}
	taps and then feed it into the Deep Neural Network (see
	Fig.~\ref{fig:fig1_diagram}C). The signal $d[n]$ is delayed by $N_{\mathrm{taps}_x}/2$
	so that the DNN has time to react to pulse-like muscle artefacts arriving at $x[n]$.
	
	The output of the Deep Neural Network $y[n]$ is then used to remove the noise from $d[n]$:
	\begin{equation}
		e[n] = d[n] - y[n]
	\end{equation}
	Ideally, this is the noise-free EEG which is at the same time
	the error signal for the DNN and back-propagated in real-time.
	Learning is ``on'' (i.e. in effect) at
	\textsl{all times}, meaning, the network adjusts to the changes in the
	electrode contact as they happen.
	
	The network used for DNF is a feed-forward neural network with fully
	connected layers designed with $L=6$ layers. The number of neurons
	$I(\ell)$ per layer index $\ell$ is calculated as:
	\begin{eqnarray}
		b &=& e^{\frac{\ln N_{\mathrm{taps}_x}}{L-1}} \\
		I(\ell) &=& \lfloor \frac{N_{\mathrm{taps}_x}}{b^{\ell-1}} \rfloor \quad \textit{ where: } \quad \ell:1, \ldots, L
	\end{eqnarray}
	which guarantees that the output layer consists of exactly one neuron
	which then generates the ``remover'' $y[n]$. In our case with
	$N_{\mathrm{taps}_x} = 50$ inputs to the DNF this results in: $I=50,
	22, 10, 4, 2, 1$ neuron(s) per layer which means that the first layer is
	fully connected with the same number of neurons to the delay line and
	then the number of neurons are reduced in the form of a funnel as done in auto-encoders.
	
	The weights of the neurons were initialised to a random value in the
	range of $(0,1]$. Eq.\ref{eq:input} below shows the forward
	propagation of the outer electrode signal $x[n]$ through the
	first layer of the network:
	\begin{equation}\label{eq:input}
		a_{j}^{0}[n] = \tanh(z^{0}_{j}[n]) = \tanh\left(\sum_{k=0}^{N_{\mathrm{taps}_x}}(\omega^{0}_{kj} x[n-k] )\right)
	\end{equation}
	where $x[n-k]$ is the filtered signal from the $k^{th}$ tap of the
	delay line for the outer electrode signal
	(Fig.\ref{fig:fig1_diagram}). In contrast to deep networks performing
	classification we filter a DC-free signal. For that
	reason, there are no bias weights to keep the processing DC-free. The
	activation function is $\tanh$ because it is ideal for signal
	processing: it is linear at the origin and becomes non-linear with
	growing signal strength so that learning can self-tune the non-linear
	processing. In the frequency domain, this means the network self-tunes
	the number of harmonics it is adding to the signals and thus to the
	remover $y[n]$.
	
	Similarly, these activations propagate through the deeper layers in
	the network:
	\begin{equation}\label{eq:f_prop}
		a_{j}^{\ell}[n] = \tanh(z^{\ell}_{j}[n]) = \tanh\left(\sum_{i=0}^{I(\ell)}\omega^{\ell}_{ij} a^{\ell-1}_{i}[n]\right) \quad \textit{ where: } \quad \ell:1, \ldots, L-1
	\end{equation}
	
	Finally, in the output layer, this weighted sum results
	in the generation of the ``Remover'' signal $y[n]$:
	\begin{equation}\label{eq:output}
		y[n] = \tanh(z^{L}_{0}[n]) = \tanh\left(\underbrace{\sum_{i=0}^{I(L)}\omega^{L}_{i} a^{L-1}_{i}[n]}_{z^{L}_{0}[n]}\right)
	\end{equation}
	
	The ``Remover'' signal $y[n]$ then ideally cancels out the noise from the inner electrode $d$:
	\begin{equation}\label{eq:subtract}
		e[n] = d[n] - y[n]
	\end{equation}
	
	As explained in previous sections, the output
	of the DNF $e[n]$ is the noise-free EEG signal and is also used for
	the learning of the neural network which is done by
	error backpropagation:
	\begin{equation}\label{backprop_1}
		\delta^{L} = e[n]
	\end{equation}
	where $\delta^{L}$ is the error in the output neuron which is then backpropagated.
	For deeper layers this is defined through the back-propagation as:
	\begin{equation}\label{backprop_2}
		\delta^{\ell}_{j} = \sum_{k=1}^{K}(w^{\ell +1}_{jk} \delta_{k}^{\ell+1})
		\cdot \tanh ' (z^{\ell}_{j}) \quad \textit{ where: } \quad \ell:L-1, \ldots, 0
              \end{equation}
              Remember that we keep the weighted sum $z^{\ell}_{j}$ well below one so that the derivative $\tanh ' (z^{\ell}_{j})$
              stays close to one preventing vanishing gradients.
	
	The changes in weights that cause the
	optimum reduction in noise are dictated by gradient
	descent rule:
	\begin{equation}\label{learning_rule}
		\Delta\omega^{\ell}_{ij}=\eta a^{\ell-1}_{i} \cdot \delta^{\ell}_{j} 
	\end{equation}
	where $\eta$ is the learning rate.
	
	It's important to note that the effective learning rate
	$\eta_\mathrm{e}$ directly scales with the amplitude of the
	noise reference $x[n]$:
	\begin{equation}
		\eta a^{\ell-1}_{i} \cdot \delta^{\ell}_{j} \equiv \underbrace{\eta x[n]}_{\eta_\mathrm{e}} \cdot e[n]
		\label{effectivelearn}
	\end{equation}
	To have a constant effective learning rate one could either normalise
	the noise reference $x[n]$ or adjust the learning rate dynamically if
	the average amplitude of $x[n]$ is changing. In this work, we directly
	set the learning rates to accommodate the two different noise reference
	amplitudes of $x[n]$ for the P300 task ($\eta=10$) and the jaw muscle task
	($\eta=2.5$) so that the effective learning rates were the same between
	the two tasks. The above equation also shows that learning converges
	when the correlation between the noise reference $x[n]$ and the error
	signal $e[n]$ weakens, meaning no frequency components of the
	noise present in the outer electrode signal remain in the output
	of the DNF filter and thus the noise has been removed.
	
	\subsection*{Calculating the signal-to-noise ratio}
	The \textsl{signal} from the inner electrode (Eq.~\ref{inner}) is a
	mix of baseline EEG, EMG and the consciously created EEG signal
	$c[n]$. To have a realistic estimate of $c[n]$ we use the
	power of the primary peak of the P300 evoked potential. To reduce the
	noise of the peak we took the median power between $300~\mathrm{ms}$ and
	$500~\mathrm{ms}$ which takes into account the $100~\mathrm{ms}$ latency of the wireless
	transmission between the ADC and the P300 software. This means that in
	terms of the power of the signal, we can think of the P300 as a pulse
	at $t=300ms$ which could be detected too, for example, by setting up a
	P300 speller. Note that the median over this time interval will
	underestimate the power slightly. However, this is deliberate because
	real-time BCI systems hardly average over 5~minutes (they do so over seconds),
	meaning they deal with much lower signal strengths for $c[n]$ and thus
	using the median filter corrects for overly optimistic signal strength.
	
	In terms of \textsl{noise}, we are interested in the power of the EMG generated
	by facial muscles and the jaw muscle but not in the low-frequency
	band such as electrooculogram (EOG) or electrode drift. To assess mainly EMG and
	underlying EEG background noise we calculated the periodogram with the
	Welch method which had a window length at the sampling rate giving the
	power density in bins of 1~Hz. The power density samples from $5~\mathrm{Hz}
	\ldots 125~\mathrm{Hz}$ were summed up given the total noise power in the frequency
	band between $5~\mathrm{Hz}$ and $125~\mathrm{Hz}$.
	
	The SNR is then calculated as:
	\begin{equation}
		\mathrm{SNR} = \frac{\mathrm{median}(v^2_{\mathrm{P300},\pm 100~\mathrm{ms}})}{\sum_{k=5~\mathrm{Hz}}^{125~\mathrm{Hz}} \mathrm{Welch}(v)[k]} \label{snrcalc}
	\end{equation}
	where $v$ can be one of the following signals: a) the inner electrode signal $d[n]$, b) the
	output $e[n]$ of the DNF, c) the output of a standard LMS-based FIR filter,
	and d) the output of the Laplace operator by directly subtracting the raw outer electrode signal
	$\tilde{d}[n]$ from the inner $\tilde{x}[n]$ one.
	
	\begin{figure}[!ht]
		\centering
		\includegraphics[width=\linewidth]{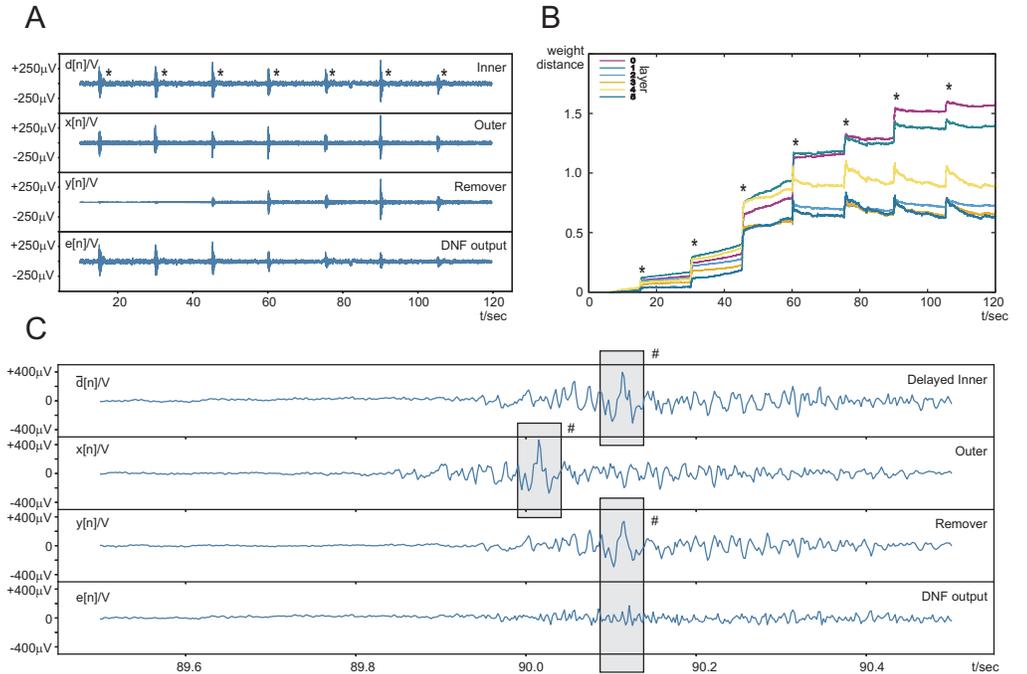}
		\caption{{\bf Signals and weight development from
		subject~10.} The subject was asked to contract
		their jaw muscles every 15~seconds. The jaw
		contractions are indicated with a ``*''. A: Four
		signal traces, namely: the inner electrode signal
		$d[n]$ which carries a mix of EEG and EMG, the outer
		electrode signal $x[n]$ which is the noise
		reference, the output of the DNN or the ``remover''
		$y[n]$, and the output of the DNF $e[n]$ which is
		both the output and the error signal. B: Weight
		development: shows the Euclidean weight distance
		from the initial weights of the 6 different layers
		over time. C: Detailed plot of the same signals
                as panel A between $89.5~\mathrm{s}$ and $90.5~\mathrm{s}$. The jaw
	        clench starts at about $89.8~\mathrm{s}$.
	\label{learning}}
	\end{figure}
	
	The recordings from the 20 subjects were then checked for valid
	EEG/EMG-signals and if deemed acceptable, processed one by one by the deep neural filter where the network had to learn from scratch (random
	re-initialisation of weights) for every subject. All parameters stayed
	the same for all subjects.
	
	\section*{Results}
	The data from the 20 subjects \cite{Bohollo2022} were examined for
	electrode failure or strong external interference. Subject~2 had a
	faulty $x[n]$ channel and subject 5 had unexplained strong artefacts
	possibly from a power surge. Thus, the results of subjects~2 and 5
	were excluded but the data from all other subjects are presented and
	analysed in this section. Before presenting the results of all
	subjects, as an instructional example, we focus on subject~10 to gain a
	deeper understanding of learning behaviour.
	Fig.~\ref{learning}A shows the progress of real-time
	learning of the DNF over a period of 2~mins for subject 10.
	``Inner''
	shows the signal $d[n]$ of the inner part of the compound
	electrode. The voluntary jaw muscle contractions every 15~seconds are
	clearly visible and indicated with a ``*''. Between the muscle
	contractions, the signal is most likely a mix of baseline EEG and lower
	amplitude involuntary facial muscle (EMG) activity. The ``Outer''
	trace shows the signal from the outer ring electrode $x[n]$ where the
	EMG bursts, caused by the jaw muscles, are clearly visible. These two
	signals, ``Inner'' and ``Outer'', are then sent into the Deep Neural
	Filter (DNF). The most important internal signal is the ``Remover''
	$y[n]$ which eliminates the noise (Eq.~\ref{eq:subtract}). The result
	of the subtraction $e[n]$ can be observed in the bottom trace ``DNF
	output''.

	Processing of the two minutes of EEG recording at
        $500~\mathrm{Hz}$ took 105~s on an Intel(R) Core(TM) i7-5600U
        CPU running at $2.60~\mathrm{GHz}$ and shows that the DNF
        filter is real-time on a general purpose processor without the
        need for special GPU hardware. The DNN has in total 6 layers
        and their weight development, related to Fig.~\ref{learning}A
        is shown in Fig.~\ref{learning}B over the two minutes. Plotted
        is the weight distance from the initial randomly initialised
        weight values. Learning is fastest during the jaw muscle
        contractions as the noise reference $x[n]$ has a higher
        amplitude and thus the effective learning rate is higher
        (Eq.~\ref{effectivelearn}) during the jaw muscle bursts but
        also continues to learn between EMG bursts at a lower
        rate. From about 60~seconds, learning has stabilised with only
        smaller adjustments to the weights till the end of the
        experiment. Because the filter acts in a closed loop
        corrective action happens where the weights shrink again after
        a jaw contraction indicating that jaw muscle recruitment and
        involuntary muscle activity cause slightly different
        correlations so that the network re-adjusts.

	Fig.~\ref{learning}C shows a zoomed-in segment of
	Fig.~\ref{learning}A between $89.5~\mathrm{s}$ and $90.5~\mathrm{s}$ at the onset of a jaw
	clench at about $89.8~\mathrm{s}$. To see how the
	removal process works the first trace $\bar{d}[n]$ shows the
	delayed version of the inner electrode signal (see
	Fig.~\ref{fig:fig1_diagram}). The noise reference $x[n]$ from
	the outer electrode is shown as it is fed into the DNF and
	then enters its tapped delay line. The DNF then creates the
	remover signal $y[n]$ which then cancels out noise in
	$\bar{d}[n]$ which is diminished at the output of the
	DNF $e[n]$ in the bottom trace which is also the error signal
	for training. As a detailed example of how the removal process works
	the section marked with the ``\#'' has been chosen. Remember that
	the DNF removes anything which is present in both the contaminated
	signal $d[n]$ and the noise reference $x[n]$. This can clearly be
	seen that the large peak is present in both the contaminated
	signal and the noise reference. Thus, the DNF learns to remove
	this peak and leaves the rest of the signal intact. Note that $e[n]$
	is also the error signal which is no
	longer correlated with the noise reference $x[n]$ which averages
	out in the learning rule Eq.~\ref{learning_rule} and consequently, the
	weights stabilise which is the case at about 90~s into learning
	(Fig.~\ref{learning}B).
	
	\begin{figure}[!ht]
		\centering
		\includegraphics[width=\linewidth]{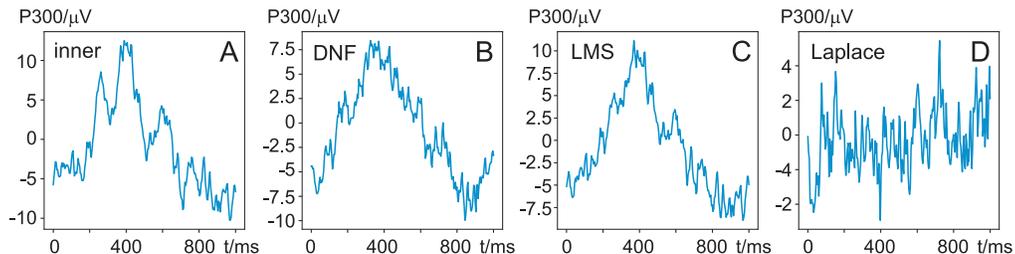}
		\caption{{\bf P300 averages from subject 10.} While looking at a
			chequerboard that inverted every second, the
			subject was presented with oddball stimuli every 7sec-13sec with a random pattern. The
			recording was 5~minutes long. A: the event-triggered
			average from the inner electrode $d[n]$,
			B: the event-triggered average from the output
			$e[n]$ of the DNF, C: the output from the LMS filter
			(adaptive FIR filter) and D: from the Laplace
			filter: $\tilde{d}[n] - \tilde{x}[n]$ with DC and
			50~Hz removed after the subtraction operation.
		\label{p300}}
	\end{figure}
	
	To calculate the SNR, the
	\textsl{power of the signal} and \textsl{the power of the
	noise} (see Eq~\ref{snrcalc}) have to be calculated separately.
	First, we focus on the
	\textsl{signal power}. As outlined above, the signal power is
	estimated by calculating the power of the primary P300 peak,
	measured during experimental session~\ref{sess2}. Note that
	there is no need to send the EEG containing the P300 through
	the DNF as the event-related averaging eliminates the EMG
	noise. However, as a sanity check we inspected the P300 peaks
	before and after noise reduction, this is shown in
	Fig.~\ref{p300} for subject~10. P300 has a low frequency by
	nature and with the DNF and LMS removing the higher EMG
	frequencies, one expects that the shape of the P300 is not
	substantially altered which is confirmed by comparing the
	unfiltered Fig.~\ref{p300}A and filtered P300
	Fig.~\ref{p300}B, C. The original EEG, the DNF output and the
	LMS filter all yielding clearly identifiable peaks and their
	squared values represent the signal power. Comparing the P300
	from the original electrode signal (A) with that of the DNF
	(B) output the P300 peak of the DNF is reduced by
	approximately $1/4$ (from $10~\mu\mathrm{V}$ to $7.5~\mu\mathrm{V}$) while the
	LMS filter causes virtually no reduction. This means that the
	DNF filter needs to reduce the noise even more than the LMS to
	achieve an overall SNR improvement, as the DNF diminishes the
	P300 peak.
	However, this is expected as there is certainly
	crosstalk between the inner electrode and the outer ring
	electrode (Eq.~\ref{outer}) where EEG from the
	inner electrode is also partially present at the outer electrode. Since the
	DNF removes anything which is present in both the noise
	reference $x[n]$ and its input signal $d[n]$ it will treat the
	$\alpha > 0$ crosstalk of the EEG signal at the outer electrode as noise and
	consequently reduces the amplitude of the noise-free EEG at its output.
	Finally, it is evident that the Laplace operator completely
	removes the P300 peak effectively rendering the SNR calculations
	for a pure Laplace operator impossible. Having calculated
	the \textsl{signal} power for the SNR, we move on to consider
	the \textsl{noise} power.

	\begin{figure}[!ht]
		\centering
		\includegraphics[width=\linewidth]{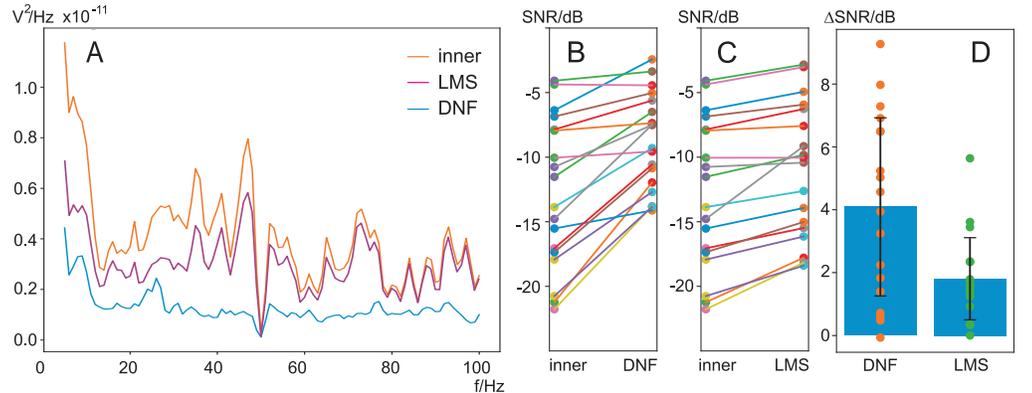}
		\caption{{\bf Noise density and SNR calculations.} A:
			Noise power density in bins of 1~Hz at the inner electrode $d[n]$, the output of the DNF $e[n]$, and the output of the standard
			LMS-based adaptive FIR filter. B: SNR in dB calculated with
			Eq.~\ref{snrcalc} at the inner electrode $d[n]$ and the output
			$e[n]$ of the DNF for every subject. C: SNR in dB calculated with
			Eq.~\ref{snrcalc} for the standard LMS-based adaptive FIR filter for
			every subject. D: The SNR differences from C) and D) for DNF
			($\Delta \mathrm{SNR}_\mathrm{DNF}=4.1 \pm 2.8$~dB) and LMS-based FIR filter ($\Delta \mathrm{SNR}_\mathrm{LMS}=1.8 \pm 1.3$~dB).
			\label{stats_noise}}
	\end{figure}
	
	In this section we calculate the \textsl{noise power} according to
	Eq.~\ref{snrcalc}. Fig.~\ref{stats_noise}A shows the power spectral
	density of the signal from the inner electrode $d[n]$ and the output
	from both the DNF and a standard LMS-based adaptive FIR filter. The
	DNF filter achieves a nearly flat reduction of the noise to about
	$0.1\cdot 10^{-11}~V^2/\mathrm{Hz}$ for frequencies above 10~Hz while
	the original noise from the inner electrode $d[n]$ fluctuates widely
	between $0.2\cdot 10^{-11} \ldots 0.8\cdot
	10^{-11}~V^2/\mathrm{Hz}$. The FIR filter tuned by LMS, being a linear
	filter with just one layer, also achieves a noise reduction but falls
	short by simply reducing the spectral components in a nearly
	proportional way and is not able to eliminate the noise peaks, for
	example at 35~Hz, 40~Hz or 45~Hz, but only reducing them. Given that
        both the DNF and the FIR filter tuned by LMS receive the same input
        signals the smooth frequency spectrum of the DNF output is clearly a distinctive
        feature of this filter.
	
	The individual SNR changes between the different subjects are shown in
	panel B and C for DNF and LMS filters, respectively. It is evident
	that the worst SNR at $-20dB$ can be improved most where strong EMG
	bursts from the jaw muscles are eliminated as shown in
	Fig.~\ref{learning}. For some subjects the improvement has been
	marginal and this might be due to poor electrode contact and thus
	little correlation between the inner and outer electrodes.
	
	To test if the noise reduction has been statistically significant,
	we calculated the SNR for every subject before and after filtering (in dB) to
	obtain the SNR improvement:
	\begin{equation}
		\Delta\mathrm{SNR} = \mathrm{SNR}_\mathrm{inner} - \mathrm{SNR}_\mathrm{DNF/LMS} 
	\end{equation}
	Fig.~\ref{stats_noise}D shows the SNR improvements for both the DNF
	and the LMS filter. Both our new DNF ($p=0.000013$) and a LMS-tuned
	adaptive FIR filter ($p=0.000192$) significantly improved the SNR but
	the DNF is significantly better than the LMS filter
	($p=0.000026$).
	
	\section*{Discussion}
	The least mean squares (LMS) technique to reduce noise in signals is
	well established \cite{Widrow1975}, where an FIR filter is trained to
	reduce the noise in a signal \cite{Hayes1996} with the help of one or
	more reference signals. This has been shown to be effective against
	EOG by using as a reference for both the horizontal and vertical EOG to
	remove the artefacts from an EEG \cite{he2004removal} but requires
	additional conventional electrodes placed above/below and left/right
	of the eyes. There have been various approaches to using neural
	networks to generate the signal (called here ``remover'') which is
	used to eliminate the artefacts in the EEG signal
	\cite{ISLAM2016}. While we use a standard encoder based deep net with
	a non-linear activation function, others used radial basis functions
	\cite{Mateo2013} or functional link neural networks (FLNN) to
	generate non-linear decision boundaries with non-linear functional
	expansion \cite{JAFARIFARMAND2013}. Even more computationally
	expensive is an approach where the shortcomings of the FLNN are
	reduced with the help of an adaptive, neural, fuzzy inference system
	\cite{HU2015}. In contrast, our deep network operates as a standard
	deep net and off-the-shelf optimised architectures are widely
	available. In particular, encoder structures are very popular across
	application domains and are readily available, for example, audio
	\cite{Braun2021}. Thus, in terms of computational cost not only the
	standard encoder architecture is beneficial because of its wide
	availability but also makes it possible to directly use deep learning
	optimised hardware such as GPUs to perform the computations.
	
	Traditionally, deep learning is a classifier and has been used to
	detect EEG artefacts with high accuracy of up to 90\%
	\cite{Craik2019,Webb2021,Bahador2020} but not to remove the artefacts
	from the EEG. Deep learning can also assist ICA-based algorithms
	\cite{Makeig1995,Delorme2007} to identify the principal components
	which contain the EMG noise \cite{Lee2020}. Direct removal of EMG
	noise has been investigated in the following network structures: fully
	connected neural networks, simple convolutional networks, complex
	convolutional networks, recurrent neural networks \cite{Zhang2021}
	and a new encoder/decoder-based architecture called DeepSeparator
	\cite{Yu2022}. Only the fully connected network, the recurrent neural
	network and the DeepSeparator were stable during EMG removal. All
	these networks received the entire time series, outputted the entire
	time series, were trained offline and are thus not real-time. In
	contrast, our DNF performs continuous real-time training and
	filtering at the same time. These networks were not trained by an
	error between reference noise and the output of the filter but by an
	error between a clean EEG and the filter output
	\cite{YANG2018,nguyen2012eog} which also served as the performance
	measure. Since clean EEGs are not readily available, they were, for
	example, generated with ICA from noisy EEGs \cite{Zhang2021}. The
	improvement of SNR before and after filtering was not stated by Zhang et al.
	\cite{Zhang2021} but the error between clean EEG and filter
	output settles at about 10\%. Overall, as Urig\"uen et al.\cite{Urig_en_2015}
	noted, most EEG noise reduction studies are based only on synthetic
	signals and most only visually analyse their results.
	
	With standard deep learning approaches \cite{Zhang2021,Yu2022} where
	learning and filtering are done separately, there is always the risk of
	overfitting \cite{Geman1992}. However, here learning is always ``on''
	which means that the DNF is constantly adapting to new signals and
	noise contingencies. The learning rate of the DNF rather determines
	how quickly it adapts where a high learning rate could lead to
	``temporary overfitting'' in particular on to large one-off artefacts
	whereas a low learning rate could not adapt to changing signal and
	noise contingencies.
	
	The mechanical EEG electrode design is as old as the first EEG
	recordings \cite{umlauf1948} and the standard Ag/AgCl cup electrodes
	have been the main staple of EEG recordings ever since
	\cite{McAdams2006}. A major concern has always been the resistance
	between electrode and skin \cite{SCHWAB1953} which has an impact on
	the SNR of the EEG. The electrode resistance has
	become even more of a concern with the advent of BCI and consumer EEG
	headbands which favour dry electrodes \cite{Guger2012}. Besides active
	electrodes \cite{Xu2017} novel electrode designs promise to help
	reduce the electrode resistance
	\cite{krachunov_casson_2016,Velcescu2019} in particular by using
	spring contact probes \cite{Viswam2015,Lun-De2019}. However, these
	electrode designs only improve the SNR by a better
	skin/electrode contact but do not take into account the spatial
	distribution of signals versus noise which calls for compound
	electrodes.
	
	The spatial distribution of electrodes has been in particular
	investigated with the rise of brain-computer interfaces (BCI) where
	often the user is actively using their muscles and thus creating a
	large amount of both EMG and movement artefacts \cite{wolpaw1997}. It
	could be shown that the central average reference (CAR) and both small
	and large Laplacian montages \cite{besio2006} improve the SNR.
	This has been shown for Electrocardiogram (ECG)
	\cite{garcia-casado2019} by removing movement artefacts and for EEG
	\cite{aghaei-lasboo2020}. Common to all approaches is the
	approximation and optimisation of a 2D spatial Laplace operator
	\cite{besio2006,makeyev_ding_besio_2016,makeyev_2018}. The more rings
	are employed at an optimal spacing the more efficient the operator
	will be. The calculation of the Laplacian is usually performed by the
	electrical summation of the EEG sources under each ring, digitisation and subtraction from each other. However, this assumes that every ring
	can perform a perfect analogue spatial averaging operation which is
	not the case in practice as electrode impedances will be inhomogeneous
	and changing over time. The analogue averaging over a ring can be
	overcome by measuring from a large number of electrodes from an EEG
	cap and then approximating the Laplace purely in software
	\cite{Fitzgibbon2015} - but this is computationally expensive and if
	using a standard EEG cap, it has its limitations in spatial
	resolution. On the other hand, the above discussed concentric ring
	electrode is the most feasible and practical hardware design
	\cite{besio2006}, however, has the drawback of assuming perfect
	recording conditions that are only present in ideal biophysical models but not in
	real setups. To overcome the shortcomings of hardwired computations
	based on ideal models we use an adaptive algorithm to account for the
	imperfect nature of the electrodes and the dynamic changes of
	electrode resistance over time, in particular when using dry
	electrodes. By high-pass filtering the noise reference (i.e outer
	electrode) we can direct the learning algorithm towards the noise it
	should focus on which here was EMG noise.
	
	A particular area of concern is the choice of adequate conductive
	electrode material. Bio-electrodes are in contact with the body and
	will, in turn, be exposed to biological electrolytes which can, over
	time, cause oxidation of the electrode and degrade the electrode's
	quality \cite{Lopez2014,Manjakkal2019,Manjakkal2020}. It can be
	concluded that precious metals are the obvious choice for conductive
	material and many EEG electrodes utilise them to provide electrode
	conductivity \cite{Mathewson2017, Flumeri2019}. Due to the cost of
	such metals, a superficial, thin coating is usually applied to a
	cheaper backing material \cite{Gorecka2019, Velcescu2019}, to provide
	high conductivity, good chemical stability and structural support for
	the electrode, simultaneously minimising the cost
	\cite{Manjakkal2018}. The conductive layer selected for the design
	discussed in this paper was also Ag/AgCl and was selected due to its
	high conductivity \cite{Velcescu2019}, chemical and electrical
	stability \cite{Lopez2014} and relative manufacturing simplicity as it can
	be printed as an ink \cite{Velcescu2019, Kalevo2020}.
	
\section*{Conclusion}
	To our knowledge, we are the first to perform simultaneous
	learning and noise reduction in real-time with a deep neural
	network without the classical sequential process of
	training first and then filtering. Specifically for removing EMG
	from EEG we have developed a novel electrode which in
	conjunction with the real-time deep learning algorithm
	implements a constantly adapting spatial Laplace filter. As a
	proof of concept, we have used data of 20 subjects performing a
	jaw-clench to produce easily identifiable EMG signals. Future
	research will focus on more realistic scenarios of EMG noise,
	for example playing a video game or performing a manual task
	where noise levels change dynamically which requires possibly
	an adaptive learning rate as used by variable step size LMS
	filters \cite{Kwong1992}. We will also investigate other
	symmetrical activation functions suitable for signal
	processing which are less computationally expensive, yield
	faster convergence and are robust against vanishing
	gradients. Generally, the DNF is also applicable to other
	domains such as noise cancelling headphones
        and will be addressed in the future.

\section*{Funding Statement}
	This work was partly supported by Engineering and Physical
	Sciences Research Council (EPSRC) through Engineering
	Fellowship for Growth - neuPRINTSKIN (EP/R029644/1). The
	funders had no role in study design, data collection and
	analysis, decision to publish, or preparation of the
	manuscript.
	
	\section*{Data Availability}
	The authors confirm that all data underlying the findings are fully available without restriction.
	All relevant data and code are referenced in the text and have citable DOIs.
	
	\section*{Conflict of Interest Statement}
	B.P. is CEO of Glasgow Neuro LTD which manufactures the Attys DAQ board.
	This does not alter our adherence to PLOS ONE policies on sharing data and materials.

\section*{Supporting information}

\paragraph*{S1 Appendix.}
\label{S1_Appendix}
{\bf DNF filtering with simulated EEG and EMG.}
	
\appendix

\section*{DNF filtering with simulated EEG and EMG}
For the simulation, the pure EEG signal and the pure EMG noise are both generated artificially. For this, we use two uncorrelated random Gaussian noise sources for both EEG and EMG, filter them
appropriately and then add them together. This has the advantage that
the SNR can directly be calculated by cross-correlating the
artificially generated pure EEG $c[n]$ with the output of the DNF
$e[n]$ to obtain the pure EEG amplitude after DNF filtering without resorting
to P300 for pure signal:
\begin{equation}
  c_\textrm{amplitude} = \max_{m = -N \ldots N} \sqrt{\frac{1}{N} \left| \sum_{n=0}^{N-1} c[n-m] \cdot e[n] \right|}
\end{equation}
where $N$ is the number of samples of the recording and $m$
an index which shifts the signals $e[n]$ and $c[n]$ against each other
until they reach a maximum resulting in the amplitude
$c_\textrm{amplitude}$ of the pure EEG at the output
of the DNF. The simulation code uses the same DNF filter code
\cite{bernd_porr_2022_7100537} as for the real EEG data. Only the code
segment that generates the simulated EEG, EMG, and the analysis using
cross-correlation instead of P300 has been adapted but otherwise kept
identical.

	\begin{figure}[!ht]
		\centering
		\includegraphics[width=\linewidth]{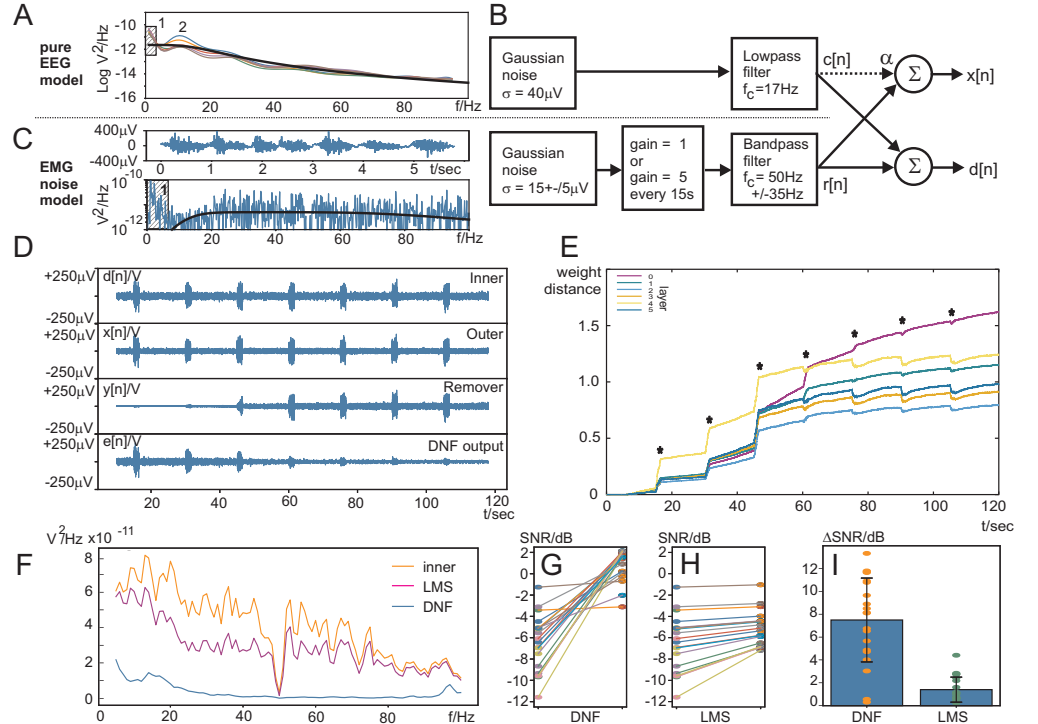}
		\caption{{\bf Simulation results.} A: Pure EEG data
                  from Whitam et al. 2007 (thin, coloured lines) and
                  2nd order Butterworth approximation (thick, black
                  line). B: Signal (EEG, $c[n]$) and noise (EMG,
                  $r[n]$) model to generate the inner $d[n]$ and outer
                  electrode $x[n]$ signals. $\alpha = 0.4$ simulates
                  the spillover of the EEG signal into the outer noise
                  reference electrode $x[n]$. C: Seven jaw clenches
                  and its Fourier transform. The thick black line in
                  the Fourier spectrum is the Butterworth bandpass
                  approximation. D: Results of the simulation
                  (matching Fig~\ref{learning}A in the main part of
                  the paper) showing the inner electrode signal
                  $d[n]$, the outer ring electrode $x[n]$, the remover
                  $y[n]$ and the output of the DNF $e[n]$. E: weight
                  development of the weights per layer matching
                  Fig~\ref{learning}B. The ``*'' indicate the
                  simulated jaw clenches. F: Noise power density in
                  bins of 1~Hz at the inner electrode $d[n]$, at the
                  output of the DNF $e[n]$ and at the output of the
                  standard LMS-based adaptive FIR filter (matching
                  Fig.~\ref{stats_noise}A).  G: SNR in dB at the inner
                  electrode $d[n]$ and the output $e[n]$ of the DNF
                  for 20 simulations. H: SNR in dB for the standard
                  LMS-based adaptive FIR filter.  I: The SNR
                  differences from G) and H) for DNF and the LMS-based
                  FIR filter.
                \label{fig:results_sim}}
	\end{figure}

	We first describe the pure EEG model: Data from Whitham et
	al. 2007 \cite{Whitham2007, bernd_porr_2022_7038831} of six
	EEG power spectra under complete paralysis
	(Fig.~\ref{fig:results_sim}A) has been used to create a model
	of pure EEG by filtering random Gaussian noise at a standard
	deviation of $40~\mu$V with a 2nd order Butterworth low-pass
	filter with a cutoff frequency of $f_c = 17$~Hz ignoring DC
	(Fig.~\ref{fig:results_sim}A1) and alpha activity
	(Fig.~\ref{fig:results_sim}A2) because our subjects had their
	eyes open in contrast to the paralysed ones. To keep
	the model simple, it is assumed that the low-pass filtered
	Gaussian noise is the pure EEG signal $c[n]$ without any additional
	EEG-background activity.
	
	The EMG signal model needs to reflect the spectrum of the jaw
	clench and is determined by taking the Fourier transform of 7
	chunks of EMG containing jaw clenches from subject 10
	(Fig.~\ref{fig:results_sim}C) which shows that below 20~Hz the
	power of the EMG rapidly declines \cite{Whitham2007}. When combining the jaw clenches, the low
	end at $f < 3$~Hz is ignored as it is not EMG but baseline
	shifts (Fig.~\ref{fig:results_sim}C1). Fig~\ref{fig:results_sim}B
	shows the generation of the simulated EMG: random Gaussian noise is
	generated at $15~\mu$V, then boosted every 15~sec for one second
	by a factor of 5 to simulate the jaw clench and then filtered with
	a bandpass filter with centre frequency of $f_c = 50$~Hz and
	bandwidth of $\pm 35$~Hz.
	
	Fig.~\ref{fig:results_sim}B shows how the pure EEG $c[n]$ and
	EMG noise $r[n]$ are combined to arrive at the simulated
	signals for the inner electrode $d[n]$ and outer ring
	$x[n]$. To factor in the fact that EEG spills over from the
	inner electrode to the outer ring electrode, a cross-talk coefficient of $\alpha = 0.4$ has
	been selected, as introduced in Eq.\ref{eq:alpha}. As mentioned in the main part, this spillover
	of the EEG signal reduces the signal amplitude at the output
	of the DNF, as the DNF treats anything at the outer ring
	electrode $x[n]$ as noise. $\alpha = 0.4$ was chosen to arrive
	at a similar reduction of the filtered EEG amplitude during
	the simulation from $10~\mu$V to $7.5~\mu$V as observed for
	the real data (Fig.~\ref{p300}A, B). Remember that there is no
	need to measure P300 as the pure EEG is known here through simulation. The
	final simulated electrode signals are now calculated as:
	\begin{eqnarray}
		d[n] = & r[n] + c[n] \quad & \textrm{Inner electrode: signal + noise} \label{innersim}\\
		x[n] = & r[n] + \alpha \cdot c[n] & \textrm{Outer ring electrode: noise reference} \label{outersim}
	\end{eqnarray}

	Fig.~\ref{fig:results_sim}D-I show the simulation results
	matching Fig.~\ref{learning} and Fig.~\ref{stats_noise} in the main
        part of the paper. The removal of the simulated jaw clenches in
	Fig.~\ref{fig:results_sim}D,$e[n]$ follows clearly a similar
	development of the removal of the real jaw clenches in
	Fig~\ref{learning}A: the remover $y[n]$ grows in amplitude and
	then successfully removes the jaw clench at the output $e[n]$
	of the DNF. The weight development of the simulation in
	Fig.~\ref{fig:results_sim}E follows also a similar development
	as the one with the real data in Fig~\ref{learning}B. Surprisingly,
	the real data makes the network converge faster than the
	simulation. The spectral distribution of the output of the DNF
	in Fig.~\ref{fig:results_sim}F is also smooth as in the
	real data in Fig.~\ref{stats_noise}A and similarly, the LMS
	filter performs worse than the DNF. This is also
	confirmed quantitatively by the signal-to-noise
	analysis. The SNR analysis over 20
	simulated subjects with different EMG noise amplitudes of
	$15\mu V \pm 5 \mu V$ reveals a very similar performance between
	simulation and real data. As for the real data, the signals
	with the lowest SNR benefit most and those with already high
	SNR benefit less; this is as expected.
        Both our new DNF ($p=0.000001$) and the LMS-tuned
        adaptive FIR filter ($p=0.000023$) significantly improved the SNR but
        as for the real data the DNF is significantly better than the LMS filter
        ($p=0.000001$). Overall, the simulation results
	confirm and validate those obtained with real data.



\begin{thebibliography}{10}

\bibitem{Green1985}
Green RM, Messick WJ, Ricotta JJ, Charlton MH, Satran R, McBride MM, et~al.
\newblock {Benefits, shortcomings, and costs of EEG monitoring.}
\newblock {Ann Surg}. 1985;201(6):785--92.
\newblock doi:{10.1097/00000658-198506000-00017}.

\bibitem{henry2006electroencephalography}
Henry JC.
\newblock Electroencephalography: basic principles, clinical applications, and
  related fields.
\newblock Neurology. 2006;67(11):2092--2092.

\bibitem{Britton2016}
Britton JW, Frey LC, Hopp JL, Korb P, Koubeissi MZ, Lievens WE, et~al.
\newblock Electroencephalography (EEG): An Introductory Text and Atlas of
  Normal and Abnormal Findings in Adults, Children, and Infants.
\newblock 1st ed. American Epilepsy Society; 2016.

\bibitem{fatourechi2007emg}
Fatourechi M, Bashashati A, Ward RK, Birch GE.
\newblock EMG and EOG artifacts in brain computer interface systems: A survey.
\newblock Clinical neurophysiology. 2007;118(3):480--494.

\bibitem{bell1995information}
Bell AJ, Sejnowski TJ.
\newblock An information-maximization approach to blind separation and blind
  deconvolution.
\newblock Neural computation. 1995;7(6):1129--1159.

\bibitem{Makeig1995}
Makeig S, Bell A, Jung TP, Sejnowski TJ.
\newblock Independent Component Analysis of Electroencephalographic Data.
\newblock In: Touretzky D, Mozer MC, Hasselmo M, editors. Advances in Neural
  Information Processing Systems. vol.~8. MIT Press; 1995.Available from:
  \url{https://proceedings.neurips.cc/paper/1995/file/754dda4b1ba34c6fa89716b85d68532b-Paper.pdf}.

\bibitem{McMenamin2010}
McMenamin BW, Shackman AJ, Maxwell JS, Bachhuber DRW, Koppenhaver AM, Greischar
  LL, et~al.
\newblock Validation of ICA-based myogenic artifact correction for scalp and
  source-localized EEG.
\newblock NeuroImage. 2010;49(3):2416--2432.
\newblock doi:{10.1016/j.neuroimage.2009.10.010}.

\bibitem{Fitzgibbon2007}
Fitzgibbon SP, Powers DMW, Pope KJ, Clark CR.
\newblock Removal of EEG noise and artifact using blind source separation.
\newblock Journal of clinical neurophysiology : official publication of the
  American Electroencephalographic Society. 2007;24(3):232--243.
\newblock doi:{10.1097/WNP.0b013e3180556926}.

\bibitem{Delorme2007}
Delorme A, Sejnowski T, Makeig S.
\newblock Enhanced detection of artifacts in EEG data using higher-order
  statistics and independent component analysis.
\newblock NeuroImage. 2007;34(4):1443--1449.
\newblock doi:{10.1016/j.neuroimage.2006.11.004}.

\bibitem{Ahmadi2012}
{Ahmadi} A, {Dehzangi} O, {Jafari} R.
\newblock Brain-Computer Interface Signal Processing Algorithms: A
  Computational Cost vs. Accuracy Analysis for Wearable Computers.
\newblock In: 2012 Ninth International Conference on Wearable and Implantable
  Body Sensor Networks; 2012. p. 40--45.

\bibitem{jirayucharoensak2013online}
Jirayucharoensak S, Israsena P, Pan-ngum S, Hemrungrojn S.
\newblock Online EEG artifact suppression for neurofeedback training systems.
\newblock In: The 6th 2013 Biomedical Engineering International Conference.
  IEEE; 2013. p. 1--5.

\bibitem{Jirayucharoensak2019}
Jirayucharoensak S, Israsena P, Pan-Ngum S, Hemrungrojn S, Maes M.
\newblock A game-based neurofeedback training system to enhance cognitive
  performance in healthy elderly subjects and in patients with amnestic mild
  cognitive impairment.
\newblock Clinical interventions in aging. 2019;14:347--360.
\newblock doi:{10.2147/CIA.S189047}.

\bibitem{Widrow1975}
Widrow B, Glover JR, McCool JM, Kaunitz J, Williams CS, Hearn RH, et~al.
\newblock Adaptive noise cancelling: Principles and applications.
\newblock Proceedings of the IEEE. 1975;63(12):1692--1716.
\newblock doi:{10.1109/PROC.1975.10036}.

\bibitem{kher2016adaptive}
Kher R, Gandhi R.
\newblock Adaptive filtering based artifact removal from electroencephalogram
  (EEG) signals.
\newblock In: 2016 International Conference on Communication and Signal
  Processing (ICCSP). IEEE; 2016. p. 0561--0564.

\bibitem{he2004removal}
He P, Wilson G, Russell C.
\newblock Removal of ocular artifacts from electro-encephalogram by adaptive
  filtering.
\newblock Medical and biological engineering and computing.
  2004;42(3):407--412.

\bibitem{Fitzgibbon2015}
Fitzgibbon SP, DeLosAngeles D, Lewis TW, Powers DMW, Whitham EM, Willoughby JO,
  et~al.
\newblock {Surface Laplacian of scalp electrical signals and independent
  component analysis resolve EMG contamination of electroencephalogram.}
\newblock {Int J Psychophysiol}. 2015;97(3):277--84.
\newblock doi:{10.1016/j.ijpsycho.2014.10.006}.

\bibitem{makeyev_ding_besio_2016}
Makeyev O, Ding Q, Besio WG.
\newblock Improving the accuracy of Laplacian estimation with novel multipolar
  concentric ring electrodes.
\newblock Measurement. 2016;80:44--52.
\newblock doi:{10.1016/j.measurement.2015.11.017}.

\bibitem{garcia-casado2019}
Garcia-Casado J, Ye-Lin Y, Prats-Boluda G, Makeyev O.
\newblock Evaluation of Bipolar, Tripolar, and Quadripolar Laplacian Estimates
  of Electrocardiogram via Concentric Ring Electrodes.
\newblock Sensors. 2019;19(17):3780.
\newblock doi:{10.3390/s19173780}.

\bibitem{aghaei-lasboo2020}
Aghaei-Lasboo A, Inoyama K, Fogarty AS, Kuo J, Meador KJ, Walter JJ, et~al.
\newblock Tripolar concentric EEG electrodes reduce noise.
\newblock Clinical Neurophysiology. 2020;131(1):193--198.
\newblock doi:{10.1016/j.clinph.2019.10.022}.

\bibitem{besio2006}
Besio G, Koka K, Aakula R, Dai W.
\newblock Tri-polar concentric ring electrode development for Laplacian
  electroencephalography.
\newblock IEEE Transactions on Biomedical Engineering. 2006;53(5):926--933.
\newblock doi:{10.1109/tbme.2005.863887}.

\bibitem{Rohaizad2019}
Rohaizad N, Mayorga-Martinez CC, Novotny F, Webster RD, Pumera M.
\newblock 3D-printed Ag/AgCl pseudo-reference electrodes.
\newblock Electrochemistry Communications. 2019;103:104--108.
\newblock doi:{10.1016/j.elecom.2019.05.010}.

\bibitem{Lee2008}
Lee S, Kruse J.
\newblock Biopotential Electrode Sensors in ECG/EEG/EMG Systems.
\newblock Analog Devices, Inc.; 2008.

\bibitem{Flumeri2019}
Di~Flumeri G, Arico P, Borghini G, Sciaraffa N, Di~Florio A, Babiloni F.
\newblock The Dry Revolution: Evaluation of Three Different EEG Dry Electrode
  Types in Terms of Signal Spectral Features, Mental States Classification and
  Usability.
\newblock Sensors. 2019;19(6):1365.
\newblock doi:{10.3390/s19061365}.

\bibitem{Lopez2014}
Lopez-Gordo M, Sanchez-Morillo D, Valle F.
\newblock Dry EEG Electrodes.
\newblock Sensors. 2014;14(7):12847--12870.
\newblock doi:{10.3390/s140712847}.

\bibitem{Suarez2018}
Suarez-Perez A, Gabriel G, Rebollo B, Illa X, Guimerà-Brunet A,
  Hernández-Ferrer J, et~al.
\newblock {Quantification of Signal-to-Noise Ratio in Cerebral Cortex
  Recordings Using Flexible MEAs With Co-localized Platinum Black, Carbon
  Nanotubes, and Gold Electrodes.}
\newblock {Frontiers in Neuroscience}. 2018;12(1).
\newblock doi:{10.3389/fnins.2018.00862}.

\bibitem{TANKISI2020243}
Tankisi H, Burke D, Cui L, {de Carvalho} M, Kuwabara S, Nandedkar SD, et~al.
\newblock Standards of instrumentation of EMG.
\newblock Clinical Neurophysiology. 2020;131(1):243--258.
\newblock doi:{https://doi.org/10.1016/j.clinph.2019.07.025}.

\bibitem{Bohollo2022}
Munoz~Bohollo L, Porr B. {EEG} and {P300} database to determine the signal to
  noise ratio during a variety of realistic tasks; 2022.
\newblock Available from: \url{https://researchdata.gla.ac.uk/1258/}.

\bibitem{bernd_porr_2022_7100537}
Porr B, Daryanavard S, Cowan H, Dahiya R. {Deep Neuronal Filter: Real-time
  noise cancellation with Deep Learning}; 2022.
\newblock Available from: \url{https://doi.org/10.5281/zenodo.7100537}.

\bibitem{Whitham2007}
Whitham EM, Pope KJ, Fitzgibbon SP, Lewis T, Clark CR, Loveless S, et~al.
\newblock Scalp electrical recording during paralysis: quantitative evidence
  that EEG frequencies above 20 Hz are contaminated by EMG.
\newblock Clinical neurophysiology : official journal of the International
  Federation of Clinical Neurophysiology. 2007;118(8):1877--1888.
\newblock doi:{10.1016/j.clinph.2007.04.027}.

\bibitem{Hayes1996}
Hayes MH.
\newblock Statistical Digital Signal Processing and Modeling.
\newblock John Wiley \& Sons; 1996.

\bibitem{ISLAM2016}
Islam MK, Rastegarnia A, Yang Z.
\newblock Methods for artifact detection and removal from scalp EEG: A review.
\newblock Neurophysiologie Clinique/Clinical Neurophysiology. 2016;46(4):287 --
  305.
\newblock doi:{https://doi.org/10.1016/j.neucli.2016.07.002}.

\bibitem{Mateo2013}
{Mateo} J, {Torres} AM, {García} MA.
\newblock Eye interference reduction in electroencephalogram recordings using a
  radial basis function.
\newblock IET Signal Processing. 2013;7(7):565--576.

\bibitem{JAFARIFARMAND2013}
Jafarifarmand A, Badamchizadeh MA.
\newblock Artifacts removal in EEG signal using a new neural network enhanced
  adaptive filter.
\newblock Neurocomputing. 2013;103:222 -- 231.
\newblock doi:{https://doi.org/10.1016/j.neucom.2012.09.024}.

\bibitem{HU2015}
Hu J, sheng Wang C, Wu M, xiao Du Y, He Y, She J.
\newblock Removal of EOG and EMG artifacts from EEG using combination of
  functional link neural network and adaptive neural fuzzy inference system.
\newblock Neurocomputing. 2015;151:278 -- 287.
\newblock doi:{https://doi.org/10.1016/j.neucom.2014.09.040}.

\bibitem{Braun2021}
Braun S, Gamper H, Reddy CKA, Tashev I.
\newblock Towards Efficient Models for Real-Time Deep Noise Suppression.
\newblock In: ICASSP 2021 - 2021 IEEE International Conference on Acoustics,
  Speech and Signal Processing (ICASSP); 2021. p. 656--660.

\bibitem{Craik2019}
Craik A, He Y, Contreras-Vidal JL.
\newblock Deep learning for electroencephalogram ({EEG}) classification tasks:
  a review.
\newblock Journal of Neural Engineering. 2019;16(3):031001.
\newblock doi:{10.1088/1741-2552/ab0ab5}.

\bibitem{Webb2021}
Webb L, Kauppila M, Roberts JA, Vanhatalo S, Stevenson NJ.
\newblock Automated detection of artefacts in neonatal EEG with residual neural
  networks.
\newblock Computer Methods and Programs in Biomedicine. 2021;208:106194.
\newblock doi:{https://doi.org/10.1016/j.cmpb.2021.106194}.

\bibitem{Bahador2020}
Bahador N, Erikson K, Laurila J, Koskenkari J, Ala-Kokko T, Kortelainen J.
\newblock A Correlation-Driven Mapping For Deep Learning application in
  detecting artifacts within the {EEG}.
\newblock Journal of Neural Engineering. 2020;17(5):056018.
\newblock doi:{10.1088/1741-2552/abb5bd}.

\bibitem{Lee2020}
Lee SS, Lee K, Kang G.
\newblock EEG Artifact Removal by Bayesian Deep Learning amp; ICA.
\newblock In: 2020 42nd Annual International Conference of the IEEE Engineering
  in Medicine Biology Society (EMBC); 2020. p. 932--935.

\bibitem{Zhang2021}
Zhang H, Zhao M, Wei C, Mantini D, Li Z, Liu Q.
\newblock {EEGdenoiseNet}: a benchmark dataset for deep learning solutions of
  {EEG} denoising.
\newblock Journal of Neural Engineering. 2021;18(5):056057.
\newblock doi:{10.1088/1741-2552/ac2bf8}.

\bibitem{Yu2022}
Yu J, Li C, Lou K, Wei C, Liu Q.
\newblock Embedding decomposition for artifacts removal in {EEG} signals.
\newblock Journal of Neural Engineering. 2022;19(2):026052.
\newblock doi:{10.1088/1741-2552/ac63eb}.

\bibitem{YANG2018}
Yang B, Duan K, Fan C, Hu C, Wang J.
\newblock Automatic ocular artifacts removal in EEG using deep learning.
\newblock Biomedical Signal Processing and Control. 2018;43:148 -- 158.
\newblock doi:{https://doi.org/10.1016/j.bspc.2018.02.021}.

\bibitem{nguyen2012eog}
Nguyen HAT, Musson J, Li F, Wang W, Zhang G, Xu R, et~al.
\newblock EOG artifact removal using a wavelet neural network.
\newblock Neurocomputing. 2012;97:374--389.

\bibitem{Urig_en_2015}
Urigüen JA, Garcia-Zapirain B.
\newblock {EEG} artifact removal{\textemdash}state-of-the-art and guidelines.
\newblock Journal of Neural Engineering. 2015;12(3):031001.
\newblock doi:{10.1088/1741-2560/12/3/031001}.

\bibitem{Geman1992}
Geman S, Bienenstock E, Doursat R.
\newblock {Neural Networks and the Bias/Variance Dilemma}.
\newblock Neural Computation. 1992;4(1):1--58.
\newblock doi:{10.1162/neco.1992.4.1.1}.

\bibitem{umlauf1948}
Umlauf CW.
\newblock {A Simplified Basal Electrode for Routine EEG Use.}
\newblock {Science}. 1948;107(2770):121.
\newblock doi:{10.1126/science.107.2770.121}.

\bibitem{McAdams2006}
McAdams E.
\newblock In: Bioelectrodes. American Cancer Society; 2006.Available from:
  \url{https://onlinelibrary.wiley.com/doi/abs/10.1002/0471732877.emd013}.

\bibitem{SCHWAB1953}
Schwab RS, Chock YC.
\newblock {A circuit for checking both electrode continuity and resistance
  during EEG recording.}
\newblock {Electroencephalogr Clin Neurophysiol}. 1953;5(3):447--9.
\newblock doi:{10.1016/0013-4694(53)90089-3}.

\bibitem{Guger2012}
Guger C, Krausz G, Allison BZ, Edlinger G.
\newblock {Comparison of dry and gel based electrodes for p300 brain-computer
  interfaces.}
\newblock {Front Neurosci}. 2012;6:60.
\newblock doi:{10.3389/fnins.2012.00060}.

\bibitem{Xu2017}
Xu J, Mitra S, Hoof CV, Yazicioglu RF, Makinwa KAA.
\newblock {Active Electrodes for Wearable EEG Acquisition: Review and
  Electronics Design Methodology.}
\newblock {IEEE Rev Biomed Eng}. 2017;10:187--198.
\newblock doi:{10.1109/RBME.2017.2656388}.

\bibitem{krachunov_casson_2016}
Krachunov S, Casson A.
\newblock 3D Printed Dry EEG Electrodes.
\newblock Sensors. 2016;16(10):1635.
\newblock doi:{10.3390/s16101635}.

\bibitem{Velcescu2019}
Velcescu A, Lindley A, Cursio C, Krachunov S, Beach C, Brown CA, et~al.
\newblock {Flexible 3D-Printed EEG Electrodes.}
\newblock {Sensors (Basel)}. 2019;19(7).
\newblock doi:{10.3390/s19071650}.

\bibitem{Viswam2015}
Nathan V, Jafari R.
\newblock {Design Principles and Dynamic Front End Reconfiguration for Low
  Noise EEG Acquisition With Finger Based Dry Electrodes.}
\newblock {IEEE Trans Biomed Circuits Syst}. 2015;9(5):631--40.
\newblock doi:{10.1109/TBCAS.2015.2471080}.

\bibitem{Lun-De2019}
Liao LD, Wang IJ, Chen SF, Chang JY, Lin CT.
\newblock {Design, fabrication and experimental validation of a novel
  dry-contact sensor for measuring electroencephalography signals without skin
  preparation.}
\newblock {Sensors (Basel)}. 2011;11(6):5819--34.
\newblock doi:{10.3390/s110605819}.

\bibitem{wolpaw1997}
McFarland D, McCane L, David S, Wolpaw J.
\newblock Spatial filter selection for EEG-based communication.
\newblock Electroencephalography and Clinical Neurophysiology.
  1997;103(3):389--394.

\bibitem{makeyev_2018}
Makeyev O.
\newblock Solving the general inter-ring distances optimization problem for
  concentric ring electrodes to improve Laplacian estimation.
\newblock BioMedical Engineering OnLine. 2018;17(1).
\newblock doi:{10.1186/s12938-018-0549-6}.

\bibitem{Manjakkal2019}
Manjakkal L, Dang W, Yogeswaran N, Dahiya R.
\newblock Textile Based Potentiometric Electrochemical pH Sensor for Wearable
  Applications.
\newblock Biosensors. 2019;.

\bibitem{Manjakkal2020}
Manjakkal L, Dervin S, Dahiya R.
\newblock Flexible Potentiometric pH sensors for Wearable Systems.
\newblock RSC Advances. 2020;10:8594--8617.

\bibitem{Mathewson2017}
Mathewson KE, Harrison TJL, Kizuk SAD.
\newblock {High and dry? Comparing active dry EEG electrodes to active and
  passive wet electrodes.}
\newblock {Psychophysiology}. 2017;54(1):74--82.
\newblock doi:{10.1111/psyp.12536}.

\bibitem{Gorecka2019}
Gorecka J, Makiewicz P.
\newblock The Dependence of Electrode Impedance on the Number of Performed EEG
  Examinations.
\newblock Sensors. 2019;19(11):2608.
\newblock doi:{10.3390/s19112608}.

\bibitem{Manjakkal2018}
Manjakkal L, Sakthivel D, Dahiya R.
\newblock Flexible Printed Reference Electrodes for Electrochemical
  Applications.
\newblock Advanced Materials Technologies. 2018;3.

\bibitem{Kalevo2020}
Kalevo L, Miettinen T, Leino A, Kainulainen S, Korkalainen H, Myllymaa K,
  et~al.
\newblock Effect of Sweating on Electrode-Skin Contact Impedances and Artifacts
  in EEG Recordings With Various Screen-Printed Ag/Agcl Electrodes.
\newblock IEEE Access. 2020;8:50934--50943.
\newblock doi:{10.1109/access.2020.2977172}.

\bibitem{Kwong1992}
Kwong RH, Johnston EW.
\newblock A variable step size LMS algorithm.
\newblock IEEE Transactions on Signal Processing. 1992;40(7):1633--1642.
\newblock doi:{10.1109/78.143435}.

\bibitem{bernd_porr_2022_7038831}
Porr B. Pure EEG power during paralysis; 2022.
\newblock Available from: \url{https://doi.org/10.5281/zenodo.7038831}.

\end{thebibliography}
\end{document}